\documentclass[12pt,a4paper,english]{article}

\usepackage{amsfonts,amssymb,graphicx
}

\usepackage{amsmath}

\makeatletter

\newcommand{\be}{\begin{equation}}
\newcommand{\ee}{\end{equation}}
\newcommand{\bear}{\begin{eqnarray}}
\newcommand{\eear}{\end{eqnarray}}

\newcommand{\e}{\,\mathrm{e}\,}

\newcommand{\im}{\,\mathrm{i}\,}
\newcommand{\diff}{\mathrm{d}}

\newcommand{\C}{{\mathbb{C}}}
\newcommand{\Z}{{\mathbb{Z}}}

\newcommand{\Sp}{{\mathbb{S}}}
\newcommand{\Pro}{{\mathbb{P}}}

\newcommand{\Tr}[1]{\:{\rm Tr}\,#1}
\newcommand{\STr}[1]{\:{\rm STr}\,#1}

\begin{document}

\begin{titlepage}

\begin{flushright}
hep-th/yymmnnn\\
HWM--07--2\\
EMPG--07--02\\
\end{flushright}

\vskip 1.8cm

\begin{centering}

{\Large {\bf Worldsheet Instantons and a Null String Limit \\[3mm]
of Born-Infeld Theory}}

\vspace{1.5cm}

{\bf Gerald A. Goldin$^{a}$, Nick E. Mavromatos$^b$}
and {\bf Richard J. Szabo$^{c}$} \\

\vspace{1cm}

$^a$ {\it Departments of Mathematics and Physics, Rutgers
University,\\ Busch Campus, Piscataway, New Jersey 08854, USA}
\\ Email: {\tt geraldgoldin@dimacs.rutgers.edu}
\\[5mm]
$^b$ {\it Department of Physics, King's College London,\\ Strand,
London WC2R 2LS, UK}
\\ Email: {\tt nikolaos.mavromatos@kcl.ac.uk}
\\[5mm]
$^c$ {\it Department of Mathematics and Maxwell Institute for
  Mathematical Sciences,\\ Heriot-Watt University, Colin Maclaurin
Building,\\ Riccarton, Edinburgh EH14 4AS, UK}
\\ Email: {\tt R.J.Szabo@ma.hw.ac.uk}
\\

\end{centering}


\vspace{2.5cm}

\begin{abstract}

\noindent For a superstring theory in four spacetime dimensions, we
propose a modification of the Born-Infeld action that possesses a
well-defined tensionless limit. We interpret this as describing
the effective target space dynamics of null strings on a D3-brane.
We argue that such a modification can be induced by nonperturbative
contributions from instantons in the worldsheet $\sigma$-model
describing string propagation on the brane.

\end{abstract}

\end{titlepage}

\noindent In this paper, we consider an abelian Born-Infeld theory
of four-dimensional superstrings on spacetime-filling D$3$-branes.
We demonstrate that the usual Born-Infeld action has a consistent
modification that possesses a well-defined null string (i.e.,
tensionless) limit, for appropriate brane worldvolumes. The
suggested form of the modified action is analogous to an earlier
proposal~\cite{Goldin:2004ea} for modifying Born-Infeld
electromagnetism to permit a consistent Galilean ($c \to \infty$)
limit. In the present, fully relativistic context, we argue that such
a modification may be induced by nonperturbative effects on the null
string worldsheet, in the form of worldsheet instantons. Our analysis
provides an independent indication that nonperturbative worldsheet
effects must be included in order to give a target space
interpretation to $\sigma$-models for null strings propagating in
gauge backgrounds. We also discuss how our results can be extended to
nonabelian Born-Infeld theory describing the target space dynamics of
coincident multiple D-branes.

Study of the tensionless limit of strings and $p$-branes spans three
decades. The original concept of the null string is due to
Schild~\cite{Schild:1976vq}, who gave a classical worldsheet action
in the limit where the Regge slope $\alpha '$ of the string becomes
infinite. Subsequently it was found that upon quantising the null
string worldsheet field theory, there appears to be \emph{no
critical dimension} of spacetime~\cite{Lizzi:1986nv}, and that this
result is consistent with conformal invariance. However, the
conclusion is still a matter of debate~\cite{Gamboa:1989px}. It has
been pointed out that the issue depends crucially on the \emph{way}
in which one quantises the theory~\cite{Bozhilov:1997xq}.
Physically, the results can be interpreted as follows. The specific
operator ordering for $p=1$ is associated with viewing the null
string as a collection of particles moving under certain conditions,
where the absence of a critical dimension is expected on physical
grounds. But if one ignores the particle interpretation, and
restricts attention only to the appropriate tensionless limit of the
conformal anomaly, then nontrivial central charge terms
appear---leading, in that case, to a critical dimension. However, it
has been argued even here that the limit in which the string tension
$T \to 0$ can be taken in a way that leads to a null string
conformal algebra with a \emph{vanishing} critical
dimension~\cite{Nicolaidis:1997mx}. These ambiguities all can be
traced back to operator orderings. By carrying out BRST quantisation
for a model of $p$-branes with second class constraints, it was
shown in~\cite{Bozhilov:1997xq} that for $p
> 1$, the operator orderings which induce the
critical dimension in the $p=1$ case are forbidden, and thus impose
no restriction on the dimensionality of the target space.

The zero-tension limit of strings and D-branes attracted renewed
attention in~\cite{Ganor:1996mu}, for both physical and
mathematical reasons. New worldsheet actions and consistent
quantisation schemes have been proposed for tensionless
strings~\cite{Savvidy:2005fe}. They have been argued to be
instrumental in a variety of applications such as the AdS/CFT
correspondence~\cite{Bredthauer:2004kv} and the twistor string
formulation of supersymmetric Yang-Mills
theory~\cite{Siegel:2004dj}. The tensionless limit of gauged,
non-compact Wess-Zumino-Witten (WZW) models, which arises when the
level $k$ of the underlying Kac-Moody algebra with a formally
divergent central charge assumes a critical value (equal to the dual
Coxeter number of the gauge group), has been analysed
in~\cite{Lindstrom:2003mg,Bakas:2005sd}. One finds that the central charge of
all higher spin generators is fixed to a critical value which is not
seen by the usual Virasoro symmetry. From a physical point of view,
such limiting cases of WZW models might describe a topological phase
of high energy quantum chromodynamics~\cite{Ellis:1998us}. While the
results of~\cite{Bakas:2005sd} do not seem to support this idea, it
was pointed out in~\cite{Ellis:1998us} that nonperturbative
worldsheet configurations---such as
instantons~\cite{Yung:1995fa}---play a crucial role in consistently
yielding the above limit, as well as in the breaking of the
topological symmetry. Such configurations were ignored
in~\cite{Bakas:2005sd}, which might account for the authors'
unexpected results concerning the decoupling of worldsheet gravity
(the Liouville mode).

Here we study the tensionless limit of strings and $p$-branes from
an alternative (but equally problematic) point of view to the
worldsheet and worldvolume perspectives described above---that of
the target space dynamics of background gauge fields in which open
strings propagate. It is well known that for oriented open strings
of coupling $g_s \ll 1$, and finite tension
\begin{equation}
T := \frac{1}{2 \pi\, \alpha '} \ne 0\,, \label{tensionalpha}
\end{equation}
propagating in weak abelian gauge field backgrounds, the bosonic
part of the low-energy effective target space dynamics is described
by the Born-Infeld action~\cite{tseytlin}--\cite{Ambjorn:2000yr},
\begin{equation}
S_{\,\rm BI}^{(p+1)} = \frac{T^{(p+1)/2}}{g_s}\, \int\,
\diff^{p+1}X~ \sqrt{-\,{\det}_{p+1}\,\left(G_{\mu\nu} +  T^{-1}\,
F_{\mu\nu} \right)}\,, \label{biaction}
\end{equation}
where the inverse power of the string coupling indicates that this
term arises from the lowest order disk diagram in open string
perturbation theory. The gauge fields $A_\mu$, $\mu=0,1,\dots,p$,
with field strength $F_{\mu\nu}$, are assumed to be living on the
longitudinal $p+1$ directions of a D$p$-brane, which from a
worldsheet perspective are characterised by $\sigma$-model fields
$X^\mu$ with Neumann boundary conditions at the boundary of the
worldsheet disk. We allow for gravitational backgrounds, so that
$G_{\mu\nu}$ is an arbitrary metric on the brane, but we set the
axion, dilaton and $B$-field to zero for simplicity. If there is a critical
target space dimension $d^*$, then the remaining $d^* -p -1$ string
embedding fields $X^i$ have fixed Dirichlet boundary conditions. If
there is no critical dimension, then one may consider
four-dimensional superstrings on spacetime-filling D$3$-branes. This
is the case to which, for definiteness, we now restrict ourselves.
For the remainder of this paper, we also set the string coupling
$g_s=1$.

Now four-dimensional abelian Born-Infeld actions are particularly
easy to manipulate, as they can be expressed in terms of two
geometrical invariants~\cite{Tseytlin:1999dj} corresponding to the
Yang-Mills and instanton densities. These may be written,
\begin{equation}
I_1 = \mbox{$\frac12$}\,G^{\mu\lambda}\,G^{\nu\rho}\,F_{\mu\nu}\,
F_{\lambda\rho}\,, \quad I_2 = -\mbox{$\frac14$}\,
\epsilon^{\mu\nu\lambda\rho}\,F_{\mu\nu}\,F_{\lambda\rho} \ .
\label{invariants}
\end{equation}
Using the identity
\begin{equation}
{\det}_4\, \left(G_{\mu\nu} + T^{-1}\, F_{\mu\nu}\right) =
\det(G)\,\left( 1 + T^{-2}\,I_1  - T^{-4}\,I_2^2\right)\,,
\label{identity}
\end{equation}
which is particular to the case of four spacetime dimensions, we can
express the Born-Infeld action (\ref{biaction}) in the form \be
S_{\,\rm BI}^{(4)} = \int\, \diff^4X~\sqrt{-\det(G)}~{\cal R}
\label{limits1}\ee where \be {\cal R} = \sqrt{T^4 + T^2\,I_1 -I_2^2}
\ . \label{limits} \ee Note that with our choice of units, the
string tension $T$ is playing the role of the critical Born-Infeld
electric field strength; i.e., the maximum electric field intensity
that can be accommodated by a real-valued Born-Infeld Lagrangian
(\ref{limits}). In the tensionless limit $\alpha'\to\infty$, the
abelian Born-Infeld action (\ref{limits1}) becomes imaginary, and is
thus undefined. This instability reflects the physical fact that in
this limit, there is no string tension to hold the strings together
in any background electric field~\cite{Ambjorn:2000yr}.

There is an analogy between this situation and the situation
discussed in~\cite{Goldin:2004ea}, where the Galilean limit $c \to
\infty$ of Born-Infeld theory was considered. In that limit, the
standard Born-Infeld action vanishes. This prompted the authors
of~\cite{Goldin:2004ea} to propose an example of how to modify the
Born-Infeld action, so as to obtain a nontrivial action and
nontrivial, nonlinear constitutive equations for electromagnetism
(or, analogously, for Yang-Mills theory) in the Galilean limit. Of
course, for strings there is no \emph{Galilean} limit \emph{per se};
but one can nevertheless follow the procedure
in~\cite{Goldin:2004ea} of introducing terms that result in a
well-defined Born-Infeld action in the \emph{null string} limit.
Thus motivated, we propose the analogous modification of the abelian
Born-Infeld Lagrangian (\ref{limits}), given by, \be {\cal R}' =
\sqrt{(T^4 + \lambda^4_2) + (T^2 + \lambda^2_1)\,I_1 - I^2_2}\,\,;
\label{modification} \ee where $\lambda_1,\lambda_2$ are appropriate
dimensionful constants, dependent on $\alpha'$, but taking finite
values $\hat\lambda_1,\hat\lambda_2$ respectively in the limit
$\alpha'\to\infty$. We describe their form and origin in more detail
below. Notice that we have modified the constant term inside the
square root of the Born-Infeld Lagrangian, which otherwise tends to
zero in the $\alpha ' \to \infty$ limit. In that limit, the modified
Born-Infeld action is now given by,
\begin{equation}
{\cal R}'_\infty:=
\lim_{\alpha' \to \infty}\,{\cal R}' = \sqrt{\hat\lambda^4_2+
\hat\lambda^2_1\,I_1 - I_2^2} \ .
\label{limitBI}
\end{equation}
The constitutive equations \be {\cal G}^{\mu\nu}\,:=\,-\frac12\,
\frac{\partial{\cal R}}{\partial F_{\mu\nu}}\,=\,-G^{\mu\lambda}\,
G^{\nu\rho}\,F_{\lambda\rho}~\frac{\partial{\cal R}}{\partial I_1}-
\frac14\,\epsilon^{\mu\nu\lambda\rho}\,F_{\lambda\rho}~
\frac{\partial{\cal R}}{\partial I_2} \label{consteqs}\ee lead to
the nonlinear electromagnetic equations of motion $\nabla_\mu{\cal
G}^{\mu\nu}=0$, and they have well-defined nonzero limits given by
\be \frac{\partial{\cal R}'_\infty}{\partial I_1}=
\frac{\hat\lambda^2_1}{2{\cal R}'_\infty}\,, \quad
\frac{\partial{\cal R}'_\infty}{\partial I_2}= -\frac{I_2}{{\cal
R}'_\infty} \ . \label{consteqslimit}\ee Notice that the invariant
structure of the modified Lagrangian is identical to that of the
original one, and it thus possesses the same worldvolume symmetries.
Requiring positivity of the argument of the square root in
(\ref{limitBI}) imposes restrictions on the relative strengths of
the various terms. For relatively weak background gauge fields, the
Born-Infeld Lagrangian is real and defined; hence the modification
(\ref{modification})-(\ref{limitBI}) is the analog of the Galilean
limit modification of the standard Born-Infeld action
of~\cite{Goldin:2004ea}. This procedure is somewhat reminescent of the
approach of~\cite{Lindstrom:1997uj}, which shows how to obtain a
well-defined Born-Infeld action in the limit where the \emph{brane}
tension vanishes. The modification in this case is interpreted
physically as the splitting of the D-brane worldvolume into a
collection of tensile strings.

Next we relate this target space modification of the Born-Infeld
action to the worldsheet dynamics of the null string. The form of
the parameters $\lambda_{j}$ can be constrained by the requirement
that they do not affect perturbative string scattering amplitudes.
The latter are expressed as series in powers of $\sqrt{\alpha '}\,
p$, where $p$ is a characteristic momentum scale of the low energy
string excitations. This implies that the $\lambda_{j}$, as
functions of $T$, should vanish faster than the inverse of any
polynomial in $T$ as $T \to \infty$; i.e., $\lambda_{j}$ approaches
$0$ faster than any power of the worldsheet $\sigma$-model coupling
constant $\sqrt{\alpha'}$ approaches $0$; so that the perturbation
expansion remains intact to all orders. For example, the functional
dependence of the $\lambda_j$ on $T$ may take the form
\begin{equation}
\lambda_{j} = \hat\lambda_{j}~\e^{-T/\hat\lambda_{j}} \ .
\label{lambdas}
\end{equation}
Then we may also propose that (\ref{lambdas}) holds in the
nonperturbative regime $\alpha'\to\infty$. The issue now is how such
$1/\alpha '$ corrections can arise at the level of the worldsheet
theory. We argue that they do so through worldsheet
instantons~\cite{Dine:1987bq}, which are already
known~\cite{Ellis:1998us,Yung:1995fa} to play a role in the
tensionless limit of strings represented by two-dimensional
$\sigma$-models.

In fact, the worldsheet $\sigma$-model action describing open string
dynamics on the D3-brane is given by
\begin{equation}
S_\sigma[X] = \frac T2\,\int\, \diff^2z~G_{i\overline i}\,\left(\,
{\overline \partial}X^i\, \partial X^{\overline i} +
\partial X^i\, {\overline \partial} X^{\overline i}\,\right)+\oint\,
\diff\tau~A_\mu\,\partial_\tau X^\mu\,. \label{instanton}
\end{equation}
Here the worldsheet is a disk, whose bulk can be regarded as a
sphere $\C\Pro^1\cong\C\cup\{\infty\}$, with local complex
coordinates $z,\overline{z}\,$; and whose boundary is a circle
$\Sp^1$ with coordinate $\tau$. We have assumed that the brane
worldvolume is a K\"ahler manifold, with Euclidean complex
coordinates $X^i,X^{\overline i}$, $i=1,2$, in order to ensure
formal convergence of the $\sigma$-model path integral
\begin{equation}
{\cal Z} = \int\,{\rm D}X ~\e^{-S_\sigma[X]}\,, \label{part}
\end{equation}
where the integration is over string maps from the disk to the
four-dimensional target space. (Analytic continuation to Minkowski
time should be done at the end.) Since the D-brane in this instance
carries a spin$^c$ structure, the global Freed-Witten anomalies cancel
in (\ref{part})~\cite{Freed:1999vc}. The metric $G_{i\overline i}(X)$
associated with the image $X$ of a spherical worldsheet is expressed
in terms of the K\"ahler potential ${\cal K}(X)$ as
\begin{equation}\label{kahlermetric}
G_{i\overline i}(X) \,=\, \partial_i \,\partial_{\,\overline i}
\,{\cal K}(X) \ .
\end{equation}
To ensure conformal invariance, we further require that the
spacetime metric $G_{i\overline{i}}$ be Ricci flat. Then there are
no harmonic $(0,2)$ or $(2,0)$ forms. The K\"ahler form
$\omega_{i\overline
  i}=-\omega_{\overline{i} i} = \im G_{i\overline{i}}$ is a harmonic
$(1,1)$ form with respect to the metric $G$.

In the traditional analysis the null string amplitudes vanish,
because in an effective action approach they appear as expansion
terms of the form $(\alpha'\,)^n\,\langle V_{i_1}\cdots
V_{i_n}\rangle$, where the $V_{i_k}$ are appropriate vertex
operators, and where the amplitude is of the order $\partial^{\,2n}$
in a derivative expansion. Thus when $\alpha'\to\infty$, one has
$\langle V_{i_1}\cdots V_{i_n}\rangle\to0$, and all local
correlation functions are trivial. This means that the worldsheet
field theory defined by (\ref{instanton})-(\ref{part}) becomes
essentially a two-dimensional topological $\sigma$-model. This
topological field theory localizes onto its instanton sectors. The
instantons are holomorphic functions of the worldsheet coordinates,
\be \frac{\partial X^i}{\partial \overline{z}} = 0 \ , \ee which
implies that only one of the two terms in the bulk action of
(\ref{instanton}) will contribute. The path integral (\ref{part})
then reduces to an integral over the finite dimensional instanton
moduli space. The worldsheet zero mode integration yields the
Born-Infeld action, while the integration over nontrivial instanton
sectors will contribute the above nonperturbative $\e^{-1/\alpha'}$
corrections to the null string amplitudes to all orders. These open
string instantons can be described as holomorphic maps from
$\C\Pro^1$ to the brane worldvolume (i.e. as holomorphic curves of
genus zero with possible self-intersections), with fixed monodromies
around non-trivial one-cycles in the spacetime prescribed by the
gauge field background $A_\mu$.

For a holomorphic instanton $X_n$ wrapping $n$ times around a single
embedded $\C\Pro^1$, the bulk contribution to the action is \be
S_n:=S_\sigma[X_n]=n\,T \ , \ee while the contribution to the path
integral from winding the disk boundary $w$ times around a single
embedded $\Sp^1$ is ${\cal W}^{\,w}$, where $\cal W$ is the Wilson
line associated to the background $A_\mu$. As these latter
contributions have no effect on the $\alpha'\to\infty$ limit of the
theory, we henceforth focus on only the bulk contributions. As we
now demonstrate, the presence of these instantons cures the
instability of ordinary Born-Infeld theory in the null string limit.
At the quantum level, they act to suppress the spontaneous creation
of charged null open strings from the vacuum~\cite{Ambjorn:2000yr,
Bachas:1992bh}.

To illustrate, we describe explicitly the case where the
four-dimensional background spacetime is the complex projective
plane $\C\Pro^2$. The qualitative results which follow, however,
hold for any Ricci-flat K\"ahler four-manifold with second Betti
number $b_2=1$, and can be generalized easily to the cases of
multiple two-cycles by considering each type of instanton in turn.
In this case, the background geometry (\ref{kahlermetric}) is
provided by the Fubini-Study metric, which can be computed from the
K\"ahler potential \be {\cal
K}(X)=\log\big(1+X^i\,X^{\overline{i}}\,\big) \ .
\label{Kahlerpotential}\ee The construction of worldsheet instantons
$X_n:\C\Pro^1\to\C\Pro^2$ proceeds as follows~\cite{crawford}. Any
such holomorphic map may be defined by setting \be
X_n(z)=\big[p_0(z)\,,\,p_1(z)\,,\,p_2(z)\big]\,, \label{Xzpiz}\ee
where $z\in\C$, the triple $[u_0,u_1,u_2]$ denotes homogeneous
coordinates on $\C\Pro^2$, and the $p_j$ are polynomials with no
common zeroes. Then the topological degree of $X_n$ is the integer
\be d=\max_{j=0,1,2}\,\big\{{\rm deg}~p_j(z)\big\} \ , \ee so that
$n=n(d,r):=d-r-2$ by the Riemann-Hurwitz theorem, where $r$ is the
total ramification index of $X_n$ with $0\leq r\leq\frac32\,d-3$. The
canonical example of such an instanton is provided by the holomorphic
map $X_n(z)=[1,(z+1)^{d-r+1},z^d\,]$.

The localization of the $\sigma$-model partition function
(\ref{part}) in the tensionless limit $T\to0$ onto an integral over
the finite dimensional instanton moduli space is in this case given
by \be {\cal Z}_\infty:= \lim_{\alpha'\to\infty}\,{\cal
Z}=\sum_{d\,\in\,\mathbb{Z}}\,
\sum_{r=0}^{\frac32\,|d|-3}\,\int_{{\cal M}_{d,r}}\,
\diff\mu(m)~\e^{-S_{n(|d|,r)}} \label{partmodsp}\ee where ${\cal
M}_{d,r}$ is the stratified moduli space of holomorphic maps
$\C\Pro^1\to\C\Pro^2$ of degree $d$ and ramification index $r$, and
the moduli $m$ can be determined from the independent polynomial
coefficients in (\ref{Xzpiz}). ${\cal M}_{d,r}$ is naturally a
connected complex manifold of complex dimension
$3d-r+2$~\cite{crawford}. Unfortunately, beyond these facts the
geometry of these moduli spaces is not generally known
(see~\cite{crawford} for some low degree examples), and so it is not
possible to further specify the moduli space integration in
(\ref{partmodsp}). Therefore, in the following we will work in a
fixed instanton {\it background\/}, and drop the integral over
moduli in (\ref{partmodsp}). This is like a dilute instanton-gas
approximation, where interactions between instantons are ignored. It
will turn out to be the correct prescription for obtaining the
string version of the $\alpha ' \to\infty$ limit discussed in the
context of the Lagrangian (\ref{limitBI}).

We denote by $\langle\! \langle {\cal O}(X) \rangle\!
\rangle:=\int\,{\rm D}X'~\e^ {-S_\sigma[X'\,]}\,{\cal
O}(X)/\int\,\diff X_0~\e^ {-S_\sigma[X_0]}$ the average of any
operator ${\cal O}(X)$ in the $\sigma$-model over the non-zero modes
$X'$ with respect to the instanton action (\ref{instanton}),
normalized by the zero-instanton partition function. Ideally, we
would like to calculate the average $\langle\!\langle {\cal
R}\rangle\!\rangle $ of the Born-Infeld operator (\ref{limits}) in a
fixed gauge field background, but in general an expansion of the
square root in unmanageable. Instead we work in the fixed instanton
background just discussed, and calculate
\begin{equation}
\big\langle\!\big\langle{\cal R}^2 \big\rangle\!\big\rangle = T^4\,
\big\langle\!\big\langle \,1\, \big\rangle\!\big\rangle + T^2\,\big
\langle\!\big\langle
G^{\mu\lambda}(X)\,G^{\nu\rho}(X)\big\rangle\!\big\rangle\,
F_{\mu\nu}\,F_{\lambda\rho} - I_2^2 \ . \label{R2avg1}\end{equation}
Note that the expectation value of the topological invariant $I_2^2$
is trivial, because it does not depend on the background geometry.
Each average in (\ref{R2avg1}) induces a sum over terms $\e^{-n\,T}
$ with the appropriate nonperturbative structures. In the limit
$T\to0$, only instantons of very large degree $n\to\infty$ will
contribute, yielding the desired modification (\ref{limitBI}) of the
Born-Infeld Lagrangian.

In the fixed instanton background, one may replace the averages in
(\ref{R2avg1}) by quantum mechanical expectation values with respect
to a state $| \psi_n \rangle$ corresponding to a specific instanton
number $n\in\Z\,$; i.e., $\langle\!\langle {\cal O}
\rangle\!\rangle:=\langle \psi_n | {\cal
  O} |\psi_n \rangle $. Here $n<0$ correspond to the contributions
from anti-holomorphic maps, and in the generalization to spaces with
second Betti number $b_2>1$, $n$ would label an integer $b_2$-vector
representing multiple instanton contributions. These states define a
complete orthonormal system of vectors in the quantum Hilbert space of
the $\sigma$-model, $\langle \psi_n |  \psi_m \rangle=\delta_{nm} $,
which decomposes the Hilbert space into superselection sectors. One
has $X|\psi_n\rangle = X_{n}\,|\psi_n\rangle $, where $X_n$ denotes
the specified instanton background such as (\ref{Xzpiz}). Inserting
the sum over such a complete set of states into (\ref{R2avg1}), the
off-diagonal matrix elements of $\cal R$ vanish as they correspond to
instanton modes belonging to distinct superselection sectors, and we
obtain
\begin{equation}
\langle \psi_n | {\cal R}^2 |\psi_n \rangle = \sum_{m\in\Z}\,
\langle \psi_n | {\cal R}|\psi_m\rangle\, \langle \psi_m | {\cal
R}|\psi_n \rangle = \langle \psi_n | {\cal R}|\psi_n \rangle^2 \ .
\end{equation}
We may therefore write the instanton-averaged Born-Infeld Lagrangian
in closed form, as
\begin{equation}
\big\langle\!\big\langle{\cal R}\big\rangle\!\big\rangle = \sqrt{T^4\,
\big\langle\!\big\langle \,1\, \big\rangle\!\big\rangle + T^2\,\big
\langle\!\big\langle
G^{\mu\lambda}(X)\,G^{\nu\rho}(X)\big\rangle\!\big\rangle\,
F_{\mu\nu}\,F_{\lambda\rho} - I_2^2} \ .
\end{equation}
This leads to the identification of the parameters $\lambda_j$ in
(\ref{limitBI}) with expressions taking the form of (\ref{lambdas}).

It should be stressed, however, that (\ref{lambdas}) is merely the
simplest form that consistent nonperturbative expressions in $\alpha
'$ might take, and that a more precise calculation of the above
instanton contributions (including the moduli space integrations)
should be expected to lead to more complicated functions. It would be
interesting to carry out an analysis in this context along the lines
of~\cite{Dine:1987bq}, who show that worldsheet instanton
contributions to tree-level target space scattering amplitudes
renormalize the spacetime superpotential and can destabilize vacuum
configurations by inducing tadpole graphs for massless particles.
Nevertheless, it is encouraging that the rough computation of
worldsheet instanton effects outlined above leads to this form.

We conclude by indicating briefly how our analysis can be
generalized to nonabelian Born-Infeld theory. For nonabelian gauge
fields, the action (\ref{biaction}) is modified by including a
symmetrized trace operation (over the group indices), denoted
$\STr$, acting on the square root~\cite{Tseytlin:1999dj}. One still
has formally the four-dimensional determinant identity
(\ref{identity}) for nonabelian field strengths. But the resulting
Lagrangian is no longer expressed in terms of just two geometrical
invariants, because $\STr({\cal R})$ includes all possible orderings
of the $F_{\mu\nu}$ factors. The formal way to proceed is first to
expand the abelian Born-Infeld Lagrangian (\ref{limits}) in (even)
powers of $F$ to define spacetime tensors ${\cal
C}^{\mu_1\nu_1\cdots\mu_{2k}\nu_{2k}}$, through the formula \be
{\cal R}=\sum_{k=1}^\infty\, {\cal
C}^{\mu_1\nu_1\cdots\mu_{2k}\nu_{2k}}~F_{\mu_1\nu_1}\cdots
F_{\mu_{2k}\nu_{2k}} \ . \label{calRexpand}\ee Writing
$F_{\mu\nu}=F_{\mu\nu}^a~T_a$ with $T_a$ the orthonormal generators
of the gauge group in the fundamental representation, the nonabelian
Born-Infeld Lagrangian is then given by \be \STr({\cal
R})=\sum_{k=1}^\infty\,d_{a_1\cdots a_{2k}}\, {\cal
C}^{\mu_1\nu_1\cdots\mu_{2k}\nu_{2k}}~F^{a_1}_{\mu_1\nu_1}\cdots
F^{a_{2k}}_{\mu_{2k}\nu_{2k}}\,, \label{STrcalRexpand}\ee where the
totally symmetric tensors $d_{a_1\cdots
  a_l}:=\STr(T_{a_1}\cdots T_{a_l})$ are the invariant tensors for the
adjoint action of the corresponding Lie algebra.

For the gauge group $SU(2)$, one can explicitly sum the series
(\ref{STrcalRexpand})~\cite{Tseytlin:1999dj}. In this case the
generators satisfy $T_a\,T_b=\delta_{ab}+\im\epsilon_{abc}\,T_c\,$,
and so $d_{a_1\cdots
  a_{2k}}=2\,\delta_{\{a_1a_2}\cdots\delta_{a_{2k-1}a_{2k}\}}$. The
simple structure of the invariant tensors enables one to write the
$SU(2)$ Lagrangian in terms of {\it three} geometric invariants,
$\Tr I_1$, $(\Tr I_2)^2$, and $\Tr I_2^2\,$, as \be \STr({\cal
R}_{SU(2)})=\sqrt{T^4+T^2\,\Tr I_1-\mbox{$\frac13$}\, \big((\Tr
I_2)^2+2\Tr I_2^2\big)} \ . \label{STrcalRSU2}\ee In the abelian
case the latter two invariants in (\ref{STrcalRSU2}) coincide, and
the Lagrangian reduces to (\ref{limits}). Again in direct analogy
with the development in \cite{Goldin:2004ea}, we have in place of
(\ref{modification}) the modification of the nonabelian Born-Infeld
action given by \be \STr({\cal
R}'_{SU(2)})=\sqrt{(T^4+\lambda_2^4)+(T^2+ \lambda_1^2)\,\Tr
I_1-\mbox{$\frac13$}\, \big((\Tr I_2)^2+2\Tr I_2^2\big)}\,,
\label{STrcalRmod}\ee which has a well-defined null string limit
with the same geometric invariant structure. In particular, the
nonabelian constitutive equations \be {\cal
G}^{\mu\nu}_a:=-\frac12\,\frac{\partial\STr({\cal R})}
{\partial F_{\mu\nu}^a} \label{nonabconst}\ee and the resulting
equations of motion again have well-defined limits as $\alpha'\to
\infty$. It is clear in this case that precisely the same worldsheet
instanton mechanism as for the abelian case gives rise to the
nonperturbative parameters $\lambda_j$ in the modified Lagrangian
(\ref{STrcalRmod}). It would be interesting to extend this analysis
to higher rank gauge groups. In general there are only partial
results describing the invariant tensors $d_{a_1\cdots a_{2k}}$, and
the Lagrangians will generically be functions of many more geometric
invariants. It would also be interesting to generalize the above
results to higher dimensional manifolds, describing string propagation
in general D$p$-branes. In~\cite{Fairlie:1999zn} it is shown how
Dirac-Born-Infeld Lagrangians can be written as square roots of
quadratic forms in any dimension.

\bigskip

\noindent {\bf Acknowledgments.} \ N.M. wishes to thank I. Bakas,
and G.G. wishes to thank V.~Shtelen, for discussions. G.G.
acknowledges the support of the Leverhulme Trust (U.K.) for a
Visiting Professorship at King's College London during his 2004-2005
sabbatical leave, when this research originated. The work of N.M.
was supported in part through the Marie Curie Research and Training
Network {\it Universenet} (MRTN-CT-2006-035863). The work of R.S.
was supported in part by the Marie Curie Research and Training
Network {\it ForcesUniverse} (MRTN-CT-2004-005104).


\begin{thebibliography}{99}

\bibitem{Goldin:2004ea}
  G.~A.~Goldin and V.~M.~Shtelen,
  ``Generalizations of Yang-Mills Theory with Nonlinear Constitutive
  Equations,''
  J.\ Phys.\ A {\bf 37}, 10711 (2004)
  [arXiv:hep-th/0401093].

\bibitem{Schild:1976vq}
  A.~Schild,
  ``Classical Null Strings,''
  Phys.\ Rev.\ D {\bf 16}, 1722 (1977).

\bibitem{Lizzi:1986nv}
  F.~Lizzi, B.~Rai, G.~Sparano and A.~Srivastava,
  ``Quantization of the Null String and Absence of Critical Dimensions,''
  Phys.\ Lett.\ B {\bf 182}, 326 (1986);
A.~Karlhede and U.~Lindstrom,
  ``The Classical Bosonic String in the Zero Tension Limit,''
  Class.\ Quant.\ Grav.\  {\bf 3}, L73 (1986);
R.~Amorim and J.~Barcelos-Neto,
  ``Strings with Zero Tension,''
  Z.\ Phys.\ C {\bf 38}, 643 (1988);
 J.~Barcelos-Neto and M.~Ruiz-Altaba,
  ``Superstrings with Zero Tension,''
  Phys.\ Lett.\ B {\bf 228}, 193 (1989);
 I.~A.~Bandos and A.~A.~Zheltukhin,
  ``Hamiltonian Mechanics and Absence of Critical Dimensions for Null
  Membranes,''
  Sov.\ J.\ Nucl.\ Phys.\  {\bf 50}, 556 (1989)
  [Yad.\ Fiz.\  {\bf 50}, 893 (1989)];
 H.~J.~de Vega, I.~Giannakis and A.~Nicolaidis,
  ``String Quantization in Curved Spacetimes: Null String Approach,''
  Mod.\ Phys.\ Lett.\ A {\bf 10}, 2479 (1995)
  [arXiv:hep-th/9412081].

\bibitem{Gamboa:1989px}
  J.~Gamboa, C.~Ramirez and M.~Ruiz-Altaba,
  ``Quantum Null (Super)Strings,''
  Phys.\ Lett.\ B {\bf 225}, 335 (1989).

\bibitem{Bozhilov:1997xq}
  P.~Bozhilov,
  ``Tensionless Branes and the Null String Critical Dimension,''
  Mod.\ Phys.\ Lett.\ A {\bf 13}, 2571 (1998)
  [arXiv:hep-th/9711136].

\bibitem{Nicolaidis:1997mx}
  F.~Lizzi,
  ``The Zero Tension Limit of the Virasoro Algebra and the Central
  Extension,''
  Mod.\ Phys.\ Lett.\ A {\bf 9}, 1495 (1994)
  [arXiv:hep-th/9404148];
  A.~Nicolaidis, J.~E.~Paschalis and P.~I.~Porfyriadis,
  ``String Tension and the Generation of the Conformal Anomaly,''
  Phys.\ Rev.\ D {\bf 58}, 047901 (1998)
  [arXiv:hep-th/9702185].

\bibitem{Ganor:1996mu}
  H.~J.~De Vega and A.~Nicolaidis,
  ``Strings in Strong Gravitational Fields,''
  Phys.\ Lett.\ B {\bf 295}, 214 (1992);
  O.~J.~Ganor and A.~Hanany,
  ``Small $E_8$ Instantons and Tensionless Non-critical Strings,''
  Nucl.\ Phys.\ B {\bf 474}, 122 (1996)
  [arXiv:hep-th/9602120];
C.~O.~Lousto and N.~Sanchez,
  ``String Dynamics in Cosmological and Black Hole Backgrounds: The
  Null String Expansion,''
  Phys.\ Rev.\ D {\bf 54}, 6399 (1996)
  [arXiv:gr-qc/9605015];
O.~J.~Ganor,
  ``Six-Dimensional Tensionless Strings in the Large $N$ Limit,''
  Nucl.\ Phys.\ B {\bf 489}, 95 (1997)
  [arXiv:hep-th/9605201];
 A.~Hanany and I.~R.~Klebanov,
  ``On Tensionless Strings in 3+1 Dimensions,''
  Nucl.\ Phys.\ B {\bf 482}, 105 (1996)
  [arXiv:hep-th/9606136];
 P.~Mayr,
  ``Mirror Symmetry, ${\cal N} = 1$ Superpotentials and Tensionless
  Strings on Calabi-Yau Fourfolds,''
  Nucl.\ Phys.\ B {\bf 494}, 489 (1997)
  [arXiv:hep-th/9610162].

\bibitem{Savvidy:2005fe}
  G.~Savvidy,
``Conformal Invariant Tensionless Strings,''
  Phys.\ Lett.\ B {\bf 552}, 72 (2003);
``Tensionless Strings: Physical Fock Space and Higher Spin Fields,''
  Int.\ J.\ Mod.\ Phys.\ A {\bf 19}, 3171 (2004)
  [arXiv:hep-th/0310085];
I.~Antoniadis and G.~Savvidy,
  ``Physical Fock Space of Tensionless Strings,''
  arXiv:hep-th/0402077;
``Tensionless Strings, Correspondence with $SO(D,D)$ Sigma-Model,''
  Phys.\ Lett.\ B {\bf 615}, 285 (2005)
  [arXiv:hep-th/0502114].

\bibitem{Bredthauer:2004kv}
  D.~Mateos, T.~Mateos and P.~K.~Townsend,
  ``Supersymmetry of Tensionless Rotating Strings in ${\rm
    AdS}_5\times\Sp^5$ and Nearly-BPS Operators,''
  JHEP {\bf 0312}, 017 (2003)
  [arXiv:hep-th/0309114];
  A.~Sagnotti and M.~Tsulaia,
  ``On Higher Spins and the Tensionless Limit of String Theory,''
  Nucl.\ Phys.\  B {\bf 682}, 83 (2004)
  [arXiv:hep-th/0311257];
  A.~Bredthauer, U.~Lindstr\"om, J.~Persson and L.~Wulff,
  ``Type IIB Tensionless Superstrings in a pp-Wave Background,''
  JHEP {\bf 0402}, 051 (2004)
  [arXiv:hep-th/0401159]; G.~Bonelli,
  ``On the Boundary Gauge Dual of Closed Tensionless Free Strings in
  AdS,''
  JHEP {\bf 0411}, 059 (2004)
  [arXiv:hep-th/0407144].

\bibitem{Siegel:2004dj}
  W.~Siegel,
  ``Untwisting the Twistor Superstring,''
  arXiv:hep-th/0404255; I.~A.~Bandos, J.~A.~de Azcarraga and
  C.~Miquel-Espanya, ``Superspace Formulations of the (Super)Twistor
  String,'' JHEP {\bf 0607}, 005 (2006)
  [arXiv:hep-th/0604037].

\bibitem{Lindstrom:2003mg}
  U.~Lindstrom and M.~Zabzine,
  ``Tensionless Strings, WZW Models at Critical Level and Massless
  Higher Spin Fields,''
  Phys.\ Lett.\  B {\bf 584}, 178 (2004)
  [arXiv:hep-th/0305098].

\bibitem{Bakas:2005sd}
  I.~Bakas and C.~Sourdis,
``On the Tensionless Limit of Gauged WZW Models,''
  JHEP {\bf 0406}, 049 (2004)
  [arXiv:hep-th/0403165];
  ``Aspects of WZW Models at Critical Level,''
  Fortsch.\ Phys.\  {\bf 53}, 409 (2005)
  [arXiv:hep-th/0501127].

\bibitem{Ellis:1998us}
  J.~R.~Ellis and N.~E.~Mavromatos,
  ``High Energy {QCD} as a Topological Field Theory,''
  Eur.\ Phys.\ J.\ C {\bf 8}, 91 (1999)
  [arXiv:hep-ph/9807451].

\bibitem{Yung:1995fa}
A.~V.~Yung,
  ``The Broken Phase of the Topological $\sigma$-Model,''
  Int.\ J.\ Mod.\ Phys.\ A {\bf 10}, 1553 (1995)
  [arXiv:hep-th/9401124];
  ``Instanton Dynamics in the Broken Phase of the Topological
  $\sigma$-Model,''
  Int.\ J.\ Mod.\ Phys.\ A {\bf 11}, 951 (1996)
  [arXiv:hep-th/9502149];
  ``Worldsheet Instantons in the $2D$ Black Hole,''
  Int.\ J.\ Mod.\ Phys.\ A {\bf 9}, 591 (1994).

\bibitem{tseytlin}
E.~S.~Fradkin and A.~A.~Tseytlin,
``Nonlinear Electrodynamics from Quantized Strings,''
  Phys.\ Lett.\ B {\bf 163}, 123 (1985);
  ``Quantum String Theory Effective Action,''
  Nucl.\ Phys.\ B {\bf 261}, 1 (1985).

\bibitem{Tseytlin:1999dj}
  A.~A.~Tseytlin,
  ``Born-Infeld Action, Supersymmetry and String Theory,''
  in: {\it The Many Faces of the Superworld}, ed. M.~A.~Shifman (World
  Scientific, Singapore, 2000), p.~417
  [arXiv:hep-th/9908105].

\bibitem{Ambjorn:2000yr}
  J.~Ambj\o rn, Y.~M.~Makeenko, G.~W.~Semenoff and R.~J.~Szabo,
  ``String Theory in Electromagnetic Fields,''
  JHEP {\bf 0302}, 026 (2003)
  [arXiv:hep-th/0012092].

\bibitem{Lindstrom:1997uj}
   U.~Lindstrom and R.~von Unge,
   ``A Picture of D-Branes at Strong Coupling,''
   Phys.\ Lett.\  B {\bf 403}, 233 (1997)
   [arXiv:hep-th/9704051]; H.~Gustafsson and U.~Lindstrom,
   ``A Picture of D-Branes at Strong Coupling II: Spinning Partons,''
   Phys.\ Lett.\  B {\bf 440}, 43 (1998)
   [arXiv:hep-th/9807064].

\bibitem{Dine:1987bq} M.~Dine, N.~Seiberg, X.-G.~Wen and E.~Witten,
``Nonperturbative Effects on the String Worldsheet,''
  Nucl.\ Phys.\ B {\bf 278}, 769 (1986);
``Nonperturbative Effects on the String Worldsheet. 2,''
  Nucl.\ Phys.\ B {\bf 289}, 319 (1987).

\bibitem{Freed:1999vc}
  D.~S.~Freed and E.~Witten,
  ``Anomalies in String Theory with D-Branes,''
  Asian J. Math. {\bf 3}, 819 (1999)
  [arXiv:hep-th/9907189].

\bibitem{Bachas:1992bh}
  C.~Bachas and M.~Porrati,
  ``Pair Creation of Open Strings in an Electric Field,''
  Phys.\ Lett.\ B {\bf 296}, 77 (1992)
  [arXiv:hep-th/9209032].

\bibitem{crawford}
L.~Lemaire and J.~C.~Wood,
``On the Space of Harmonic $2$-Spheres in $\C\Pro^2$,''
Internat. J. Math. {\bf 7}, 211 (1996)
[arXiv:dg-ga/9510003];
T.~A.~Crawford,
``The Space of Harmonic Maps from the 2-Sphere to the
Complex Projective Plane,''
Canad. Math. Bull. {\bf 40}, 285 (1997)
[arXiv:dg-ga/9512004].

\bibitem{Fairlie:1999zn}
  D.~B.~Fairlie,
  ``Dirac-Born-Infeld Equations,''
  Phys.\ Lett.\  B {\bf 456}, 141 (1999)
  [arXiv:hep-th/9902204];
  ``Lagrange Brackets and $U(1)$ Fields,''
  Phys.\ Lett.\  B {\bf 484}, 333 (2000)
  [arXiv:hep-th/0005009].

\end{thebibliography}
\end{document}